# Topological insulator based spin valve devices: evidence for spin polarized transport of spin-momentum-locked topological surface states

Jifa Tian[1,2], Isaac Childres[1,2], Helin Cao[1,2], Shen Tian[1], Ireneusz Miotkowski[1], Yong P. Chen[1,2,3,*]

1. Department of Physics and Astronomy, Purdue University, West Lafayette, Indiana 47907, USA

2. Birck Nanotechnology Center, Purdue University, West Lafayette, Indiana 47907, USA

3. School of Electrical Engineering, Purdue University, West Lafayette, Indiana 47907, USA

*Email: yongchen@purdue.edu

**Abstract**

Spin-momentum helical locking is one of the most important properties of the nontrivial topological surface states (TSS) in 3D topological insulators (TI). It underlies the iconic topological protection (suppressing elastic backscattering) of TSS and is foundational to many exotic physics (eg., majorana fermions) and device applications (eg., spintronics) predicted for TIs. Based on this spin-momentum locking, a current flowing on the surface of a TI would be spin-polarized in a characteristic in-plane direction perpendicular to the current, and the spin-polarization would reverse when the current direction reverses. Observing such a spin-helical current in transport measurements is a major goal in TI research and applications. We report spin-dependent transport measurements in spin valve devices fabricated from exfoliated thin flakes of $Bi_2Se_3$ (a prototype 3D TI) with ferromagnetic (FM) Ni contacts. Applying an in-plane magnetic ($B$) field to polarize the Ni contacts along their easy axis, we observe an asymmetry in the hysteretic magnetoresistance (MR) between opposite $B$ field directions. The "polarity" of the asymmetry in MR can be reversed by reversing the direction of the DC current. The observed asymmetric MR can be understood as a spin-valve effect between the current-induced spin polarization on the TI surface (due to spin-momentum-locking of TSS) and the spin-polarized ferromagnetic contacts. Our results provide a direct transport evidence for the spin helical current in TSS.

## Introduction

Three-dimensional (3D) topological insulators (TIs) represent an interesting new class of quantum matter hosting spin helical surface states protected by time-reversal symmetry.[1-7] The nontrivial topological surface states (TSS, depicted in Fig. 1a) located inside the bulk band gap feature a characteristic spin-momentum-locking, where charge carriers of given momentum ($\vec{k}$) are spin polarized in-plane and perpendicularly "locked" to $\vec{k}$. For electrons, the spin polarization ($\vec{s}$) is along the direction of $\vec{k} \times \vec{n}$ ($\sigma^-$ helicity, governed by the left hand rule, depicted in Fig. 1b) with $\vec{n}$ being the surface normal, and holes have the opposite polarization ($-\vec{k} \times \vec{n}$, $\sigma^+$ helicity, right handed spin-momentum locking). A directional electrical current (*I*) carried by such spin-helical TSS would be automatically spin-polarized (noting the spin polarization for a given *current* direction is the same regardless whether the current is carried by electrons or holes, as electron momentum is opposite to the current direction), and its spin polarization reverses upon reversing the current direction (depicted in Fig. 1c,e), or going to the opposite surface (reversing $\vec{n}$). The spin-momentum locking of TSS is the basis of the topological protection (as a backscattering that reverses momentum would have to reverse the spin) and many other exotic physics predicted for TI (eg. majorana fermions [4,8]), and the expected helical spin-polarized transport makes TI particularly promising for spintronics device applications [4,9-12]. While the existence of the spin-momentum-locked TSS in 3D TIs has been established by spin and angle resolved photoemission spectroscopy (spin ARPES) measurements [13-19], direct demonstration of the spin-helical current (current induced spin polarization) using spin-sensitive transport measurements have been lacking till very recently [20-22], even though various different theoretical proposals have been discussed [9-12]. Previously, the spin valve effect (where a current flows through two ferromagnets (FM) of parallel magnetizations with lower resistance and antiparallel magnetizations with higher resistance) and spin valve devices have been commonly used to study spin transport in various materials (including metals, semiconductors, and graphene) [23-27]. Inspired by this, we have fabricated TI-based spin valve devices from exfoliated thin flakes of $Bi_2Se_3$ (a prototype 3D TI [28,29]) with FM (Ni) contact electrodes, and performed spin-valve measurements where the magneto resistance (MR) between two Ni contacts is monitored as a function of an in-plane magnetic (*B*) field applied to magnetize the Ni contacts along their easy axis (perpendicular to the current). We observe an *asymmetry* in the MR between the opposite limits of *large* positive/negative B field (where both FM contacts are magnetized along a common direction that reverses between the two limits). Furthermore, the "polarity" of this MR asymmetry reverses when the direction of the DC current is reversed. This effect (current-direction-reversible spin-valve MR asymmetry between opposite large B fields) has not been observed in previous spin-valve devices on other materials (where MR is symmetric between opposite large B fields), and can be interpreted as a spin-valve effect between the TI channel (which has a current-induced spin polarization via spin-momentum-locking of TSS) and both FM electrodes (whose common magnetization direction may be parallel or anti-parallel with the TI surface current spin polarization). Our results give a direct transport evidence for the spin helical current of TSS in a 3D TI $Bi_2Se_3$.

## Experimental

The high quality bulk $Bi_2Se_3$ single crystal is grown by the Bridgeman method [30-32]. Thin flakes of 10-20 nm in thickness are exfoliated from the bulk crystal using the standard "scotch tape" method [30, 33-34] and are placed on top of heavily doped Si substrates with 300 nm $SiO_2$. The FM electrodes (Ni, thickness=40nm, length ~ 3μm, width between 200nm and 800nm) crossing and contacting the TI top surface are defined by standard e-beam lithography and deposited by e-beam evaporation. These Ni electrodes are contacted further outside the $Bi_2Se_3$ flake by Au electrodes fabricated by a second e-beam lithography and evaporation. In this work, we have selected flakes of relatively narrow width (~1 μm) and performed two-terminal spin-valve measurements (resistance between two FM electrodes) using a DC bias current $I$ and an in-plane $B$ field (see Fig. 1c-f for device and measurement schematics). The voltage ($V$) difference is measured between the FM (Ni) electrodes (labeled by Ni1 and Ni2), and the magnetoresistance (MR) is defined by $R=V/I$. Hereafter, we define $+I$ ($-I$) direction as from Ni1 to Ni2 (Ni2 to Ni1) along $+x$ ($-x$) axis and the positive (negative) in-plane $B$ field points to the $+y$ ($-y$) axis indicated by the yellow arrows, respectively, as depicted in Fig. 1c. At a fixed bias current $I$, we sweep the $B$ field from a sufficiently large positive value (far exceeding the coercive fields of the Ni electrodes, so that both Ni electrodes are magnetized along +y direction, depicted in Fig. 1c) through zero and to a large negative value (both Ni electrodes magnetized along –y direction, depicted in Fig. 1d) and then sweep back again to the starting large positive $B$ field. We then reverse the direction of the bias current and repeat the above measurements (Fig. 1e,f). Results from two devices are presented below.

**Results and discussions**

Fig. 2 shows the results of magnetoresistance measurements in our spin valve device "*A*", fabricated on a 12 nm-thick exfoliated $Bi_2Se_3$ flake. The inset of Fig. 2a shows the optical image of the device, where the spacing between the two Ni electrodes is 200 nm. The measurements were made using bias current $|I|$=100 nA and at temperature $T$=0.3 K. The relatively high 2-terminal resistance (~200 kΩ) of this device is attributed to a large contact resistance likely resulted from an unclean interface between TI and contacts (eg. due to surface contaminants from fabrication process). The MR measured between -2T and +2T is shown in Fig. 2ab and a zoomed-in view (between -0.5T and +0.5T) shown in Fig. 2cd. The first set of features one may notice are some drastic resistance jumps observed at very low $B$ field (<<0.4T), where R vs $B$ is hysteretic and goes up and down several times. Such complicated features at low $B$ are quite different from the MR features observed in previously studied planar spin-valve devices on non-TI materials (where the MR typically displays a single resistance "bump" on each side of zero $B$ field depending on the $B$ field sweep direction) [23-27], and most of these features we observe remain to be better understood. They may be related to the spin valve effects between the two Ni electrodes (whose magnetization switches at such low fields) that may also involve the polarized spin of TSS in the TI channel in a way that substantially modifies the usual spin-valve behavior seen in non-TI materials. Furthermore, the formation and switching of multiple different magnetic domains in the Ni electrodes could also play some roles. However, most of these complicated features at low $B$ are not always observed in other devices we studied (see Fig 3), and are not the focus of in this paper. Our main feature in the MR that we focus on here is the more subtle *asymmetry* in the MR between large positive ($B$>0.5 T) and large negative (B<-0.5T) fields, where the magnetization $\vec{M}$ in both the FM electrodes are parallel and points along +y or – y directions for such large +$B$ or –$B$ field respectively. In Fig. 2a, the asymmetric MR manifests as a "high R" state of ~196 kΩ for $B$<-0.4T and a "low R" state of ~193 kΩ for $B$>0.4T, observed in the MR data from both $B$ field sweep directions. Such an unusual asymmetry in MR between opposite large $B$ fields is not observed in previously studied spin valve devices in non-TI materials (even spin-orbit coupled semiconductors such as InAs [24]), where the "asymptotic" MR at large B fields (when the two FM electrode share the same magnetization direction) is the same between the opposite $B$ field directions (when the two FM electrodes both reverse their common magnetization direction) [23-27]. The asymmetry in MR could be consistent with the existence of a substantial spin polarization ($\vec{S}$) in the channel that is not reversed when the *B* field is reversed. Most strikingly, we find that the "polarity" of the above asymmetry in MR is *reversed* by reversing the current direction (I~ -100 nA, as shown in Fig. 2b,d, where the "high R" state now occurs for $B > 0.4T$ and "low R" state now occurs for $B < -0.4T$), suggesting that channel spin polarization is reversed by reversing *I* (thus "locked" to the current direction). We can further define a normalized spin-valve signal β in terms of the asymmetry (difference between large +*B* and –*B*) in terms of relative MR $\Delta R/R_0 = (R(B) - R_0)/R_0$, $R_0$ being the average of the *R(B)*, as shown in the right axes of Fig. 2. For device "A", we find β is around 1-2 %.

We have also performed similar spin valve measurements in another device "B" and further studied the temperature effect, shown in Fig. 3. Device "B" is fabricated from a 20 nm-thick exfoliated $Bi_2Se_3$ flake and our measurements were performed with bias $I= \pm 1$ μA at *T* of 1.4 K and 10 K. At T=1.4K, shown in Figs. 3a,b, we see again the asymmetric MR between large +*B* and –*B* fields (with also a clear hysteresis near zero field), consistent with the spin valve effect between TSS and FM electrodes (with a normalized spin valve signal β~0.1%, much lower than that of device "A"). We note that the more complicated features at low *B* (in Fig. 2 for device "A") are not observed here in device "B" within experimental resolution. As *T* was increased to 10 K, such asymmetric in the MR is no longer observable (Figs. 3c,d) within the experimental resolution. This weakening and disappearance of spin-valve signal at elevated T may be related to increased scattering of carriers, or thermal activation of carriers from spin polarized TSS to bulk conduction bands (which may even carry opposite spin polarization from the TSS [35]).

Our observation can be qualitatively understood as a spin-valve effect between the current-induced spin polarization of TSS on the TI top surface and the spin-polarized FM contacts as depicted in Figs. 1c-f. Here, we focus on the case under a large in-plane *B* field such that the magnetization ($\vec{M}$) of the FM electrodes has the same orientation (either +y or –y) along the easy axis. Inspired by the well-known spin valve and GMR effect between two ferromagnets with antiparallel (leading to high R) or parallel (leading to low R) [23-27,36,37], one expects a high R state when the spin polarization $\vec{S}$ (whose direction is determined by the current direction according to the spin-momentum locking of TSS, as depicted in Fig.1) in the channel (top surface of TI) is antiparallel ("in disagreement") with the orientation of $\vec{M}$ (determined by the direction of *B* field) in both FM electrodes (Fig. 1de), while a low R state when $\vec{S}$ and $\vec{M}$ are parallel ("in agreement", Fig. 1cf). Reversing the direction of (large) *B* field reverses $\vec{M}$ but does not change $\vec{S}$, thus giving rise to an asymmetry in MR between large +*B* and –*B* fields. Reversing *I* reverses $\vec{S}$, thus reversing the "polarity" of the MR asymmetry. Our model suggests that the two-terminal MR of a TI-FM spin valve device is not only controlled by the magnetization of the FM electrodes, but also the applied DC bias *I*. This model further suggests that there would be a "symmetry" (expecting the same MR) if both *B* field ($\vec{M}$) and current *I* ($\vec{S}$) are *simultaneously* reversed. We note that such a symmetry (upon reversing both *B* and *I*) does indeed hold approximately for the data shown in Fig. 2 (the MR curve measured sweeping from +*B* to –*B* at *I* nearly reproduces the curve measured sweeping from –*B* to +*B* at -*I*).

Recently and during the preparation of this paper, we became aware of the work by Li et al., [20] which report a transport signature of spin-momentum-locking of TSS in MBE (molecular beam epitaxy)-grown $Bi_2Se_3$ devices (of much larger size than ours). Their measurements appear to be in the linear response regime (with a voltage signal that is proportional to the current *I*, thus a current-independent resistance, and the voltage reverses under *B* reversal). In contrast, our measurement is in the non-linear response regime (with a *resistance* signal that depends on the current and *B* directions), and cannot be described by the Onsager relationship (which states two-terminal resistance should be symmetric with *B* field) that only applies to linear-response regime. Our observed asymmetry in MR thus provides another signature of spin-helical transport in TSS. In addition, we also noted 2 other preprints reporting signatures of spin-helical TSS transport measured using spin torque [22] and spin pumping [21] techniques.

In conclusion, we have fabricated spin valve devices on exfoliated $Bi_2Se_3$ thin films and performed two-terminal spin valve (magnetoresistance, with in-plane *B* field) between two FM contact electrodes (magnetized by the *B* field). By driving a DC current, we find that the two-terminal resistance is asymmetric between large positive and negative *B* fields. The "polarity" of the asymmetry can be reversed by reversing the direction of the bias current. Furthermore, the measured resistance asymmetry decreases as temperature increases. Our observation is consistent with the spin-momentum helical locking of TSS producing a spin-polarized helical current, and opens ways to utilize such a remarkable property of TI for future applications in nanoelectronics and spintronics.

**Acknowledgements**

We acknowledge support by a joint seed grant from the Birck Nanotechnology Center at Purdue and the Midwestern Institute for Nanoelectronics Discovery (MIND) of Nanoelectronics Research Initiative (NRI), and by DARPA MESO program (Grant N66001-11-1-4107). We also acknowledge the valuable discussions from Prof. S. Datta and S. Hong at Purdue University.

**Figure Captions**

**Figure 1** Schematic of the experimental design to probe the spin helical TSS. (a) TI band structure with the bulk conduction band (BCB), topological surface states (TSS, arrows indicating the top surface spin polarization due to spin-momentum-locking), bulk valence band (BVB). (b) Schematic of the electron Fermi surface at top surface, in the $k_x$-$k_y$ plane of TSS with spin polarization showing $\sigma^-$ helicity. (c-f) Schematic of a TI-based spin valve device and a TI-FM spin valve effect between the current-induced spin polarization ($\vec{S}$) of TSS (top surface) and magnetic field-induced spin polarization (magnetization, $\vec{M}$) of FM contacts. We expect a lower resistance (R) state if the TSS spin polarization is parallel to the FM magnetization (c,f) and a higher R state if they are anti-parallel (d,e). Reversing the current direction reverses TSS spin polarization and reversing the magnetic field ($B$, assumed to be sufficiently large) reverses the magnetization of both FM contacts. The resistance is measured between the two FM contacts.

**Figure 2** Observation of TI-FM spin valve effect, indicating spin-helical current of TSS. (a,b) Magnetoresistance (MR, shown in both actual value, left axis, and relative change, right axis) measured from an exfoliated 12-nm-thick $Bi_2Se_3$ thin film (device "A", inset is the device image, the scale bar is 1 μm) with a DC current ($I=\pm 100$ nA) of both directions at $T=0.3$ K. A MR asymmetry (between large $+B$ and $-B$ fields) is clearly observed (with polarity reversed upon reversing current direction), consistent with the TI-FM spin valve effect depicted in Fig. 1 due to current induced spin polarization of spin-momentum-locked TSS; (c,d) zoomed-in view of (a,b). Arrows in the insets of (a,b) schematically show the directions of FM magnetization ($\vec{M}$) and current (I)-induced spin polarization ($\vec{S}$) of TSS.

**Figure 3** TI-FM spin valve effect measured in another device, and effect of temperature. (a-b) Asymmetric MR (shown in both actual value, left axis, and relative change, right axis) measured in a 20 nm-thick $Bi_2Se_3$ flake (device "B") at (a,b) $T=0.3$ K, exhibiting the TI-FM spin valve effect due to the spin-helical TSS, while the asymmetry is no longer observed when $T$ is raised to 10 K (c,d). Arrows in the insets of (a,b) schematically show the directions of FM magnetization ($\vec{M}$) and current (I)-induced spin polarization ($\vec{S}$) of TSS.

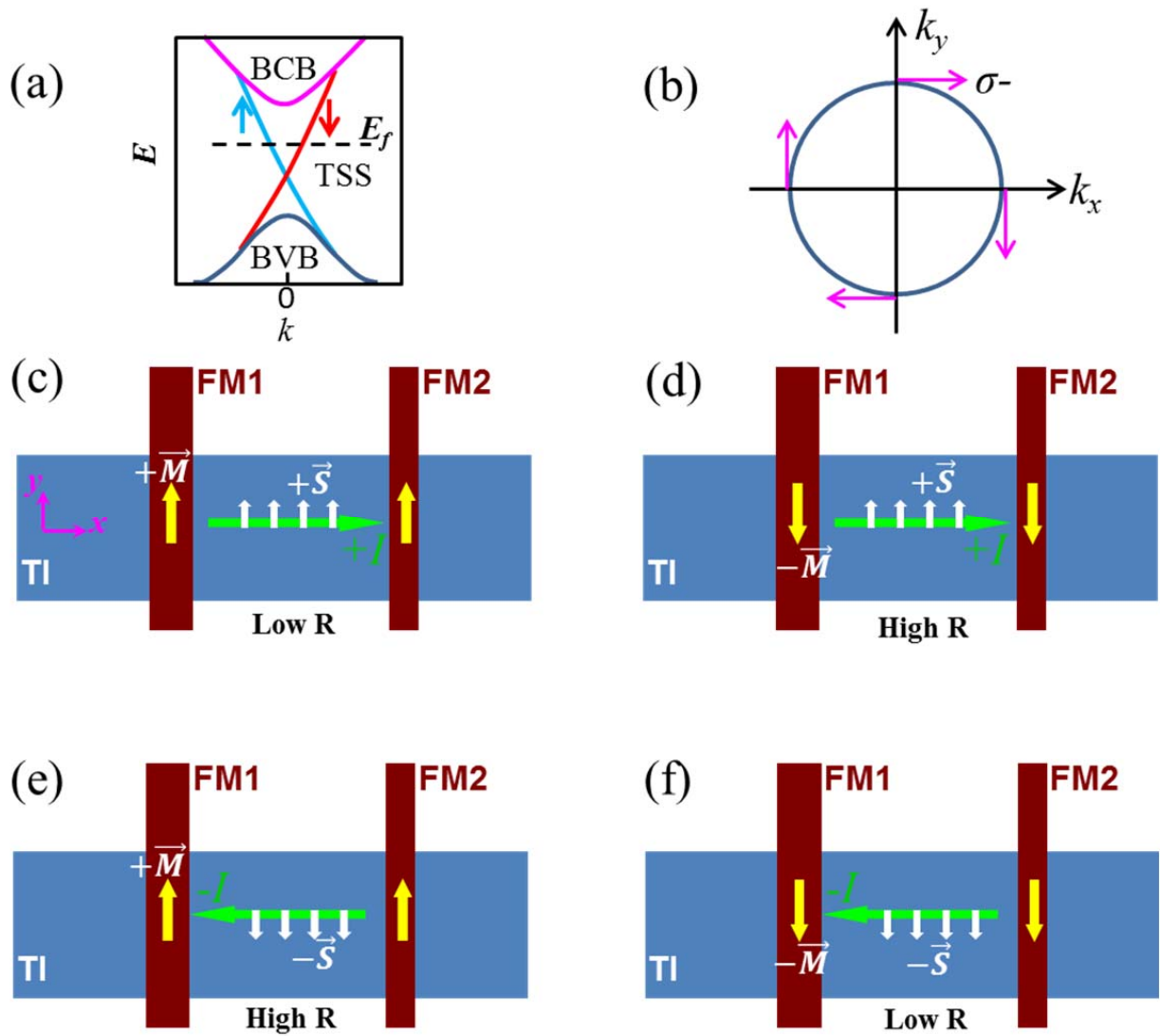

(Figure 1 by Jifa Tian et al.)

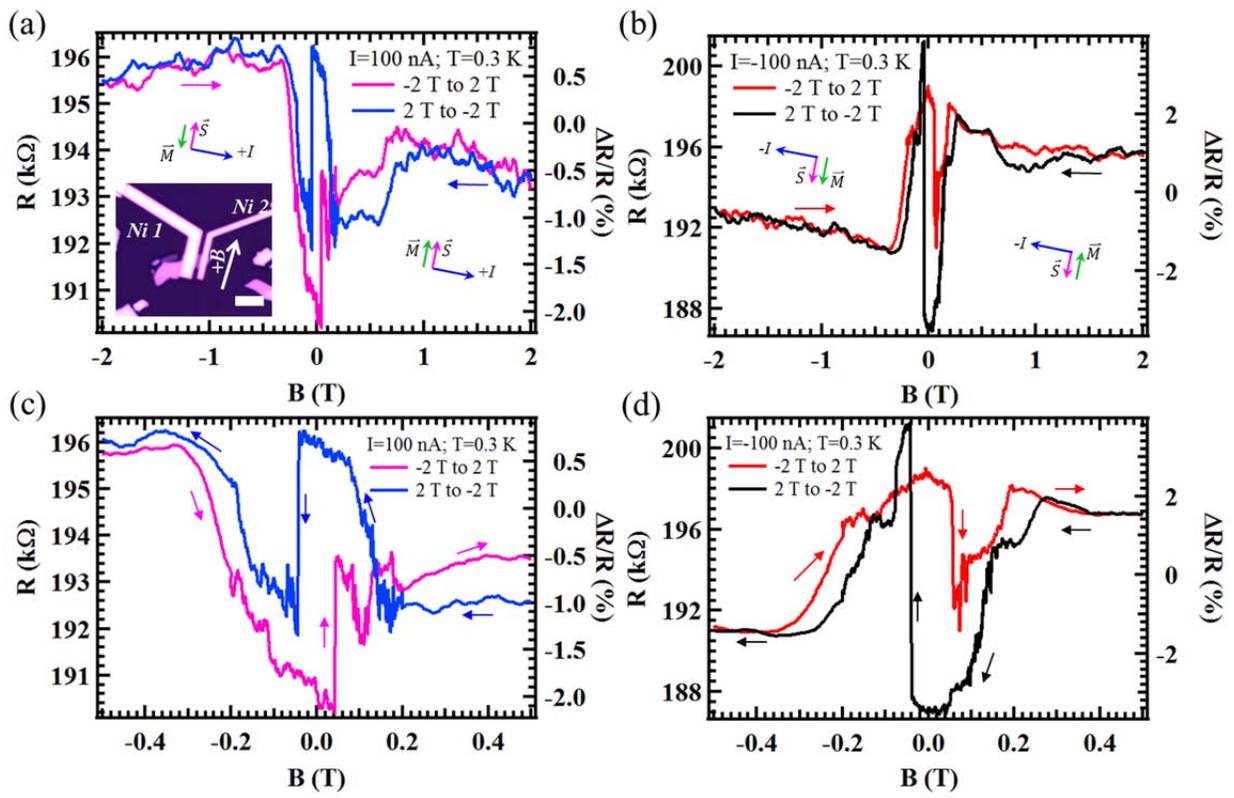

(Figure 2 by Jifa Tian et al.)

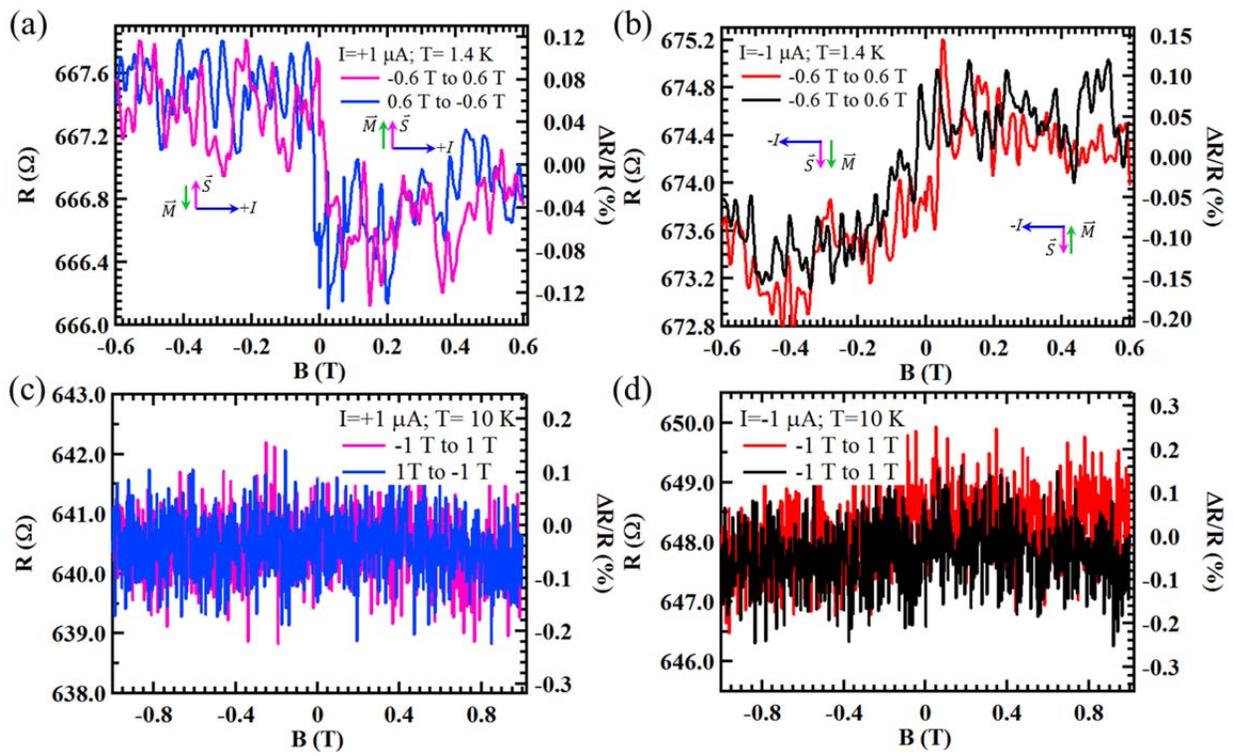

(Figure 3 by Jifa Tian et al.)